\documentstyle[12pt]{article}
\newcommand{\newsection}[1]{
\addtocounter{section}{1}
\setcounter{equation}{0}
\setcounter{subsection}{0}
\addcontentsline{toc}{section}{\protect
\numberline{\arabic{section}}{{\rm #1}}}
\vglue .6cm
\pagebreak[3]
\noindent{\bf  \thesection. #1}\nopagebreak[4]\par\vskip .3cm}
\newcommand{\newsubsection}[1]{
\addtocounter{subsection}{1}
\addcontentsline{toc}{subsection}{\protect
\numberline{\arabic{section}.\arabic{subsection}}{#1}}
\vglue .4cm
\pagebreak[3]
\noindent{\it \thesubsection. #1}\nopagebreak[4]\par\vskip .3cm}

\renewcommand{\theequation}{\thesection.\arabic{equation}}

\newcommand{\ben}{\begin{enumerate}}
\newcommand{\een}{\end{enumerate}}
\newlength{\extraspace}
\newlength{\extraspaces}
\newcounter{dummy}
\newcommand{\bc}{\begin{center}}
\newcommand{\ec}{\end{center}}
\newcommand{\be}{\begin{equation}
\addtolength{\abovedisplayskip}{\extraspaces}
\addtolength{\belowdisplayskip}{\extraspaces}
\addtolength{\abovedisplayshortskip}{\extraspace}
\addtolength{\belowdisplayshortskip}{\extraspace}}
\newcommand{\ee}{\end{equation}}

\newcommand{\ba}{\begin{eqnarray}
\addtolength{\abovedisplayskip}{\extraspaces}
\addtolength{\belowdisplayskip}{\extraspaces}
\addtolength{\abovedisplayshortskip}{\extraspace}
\addtolength{\belowdisplayshortskip}{\extraspace}}
\newcommand{\ea}{\end{eqnarray}}

\newcommand{\ban}{\begin{eqnarray*}
\addtolength{\abovedisplayskip}{\extraspaces}
\addtolength{\belowdisplayskip}{\extraspaces}
\addtolength{\abovedisplayshortskip}{\extraspace}
\addtolength{\belowdisplayshortskip}{\extraspace}}
\newcommand{\ean}{\end{eqnarray*}}
\newcommand{\baa}{                         
\addtocounter{equation}{1}
\setcounter{dummy}{\value{equation}}
\setcounter{equation}{0}
\renewcommand{\theequation}{\thesection.\arabic{dummy}\alph{equation}}
\begin{eqnarray}
\addtolength{\abovedisplayskip}{\extraspaces}
\addtolength{\belowdisplayskip}{\extraspaces}
\addtolength{\abovedisplayshortskip}{\extraspace}
\addtolength{\belowdisplayshortskip}{\extraspace}}
\newcommand{\eaa}{                                       
\end{eqnarray}
\setcounter{equation}{\value{dummy}}
\renewcommand{\theequation}{\thesection.\arabic{equation}}}

\input epsf

\newcounter{fignum}
\newcounter{tabel}

\newcounter{tabnum}
\newcounter{xxx}
\newcommand{\bl}{\begin{list}{({\it\roman{xxx}})}{\usecounter{xxx}}}
\newcommand{\el}{\end{list}}
\newcommand{\ppt}[1]{{\partial \over \partial t}}            
\newcommand{\ppx}[1]{{\partial \over \partial x}}            
\newcommand{\pqt}[1]{{\partial^2 \over \partial t^2}}            
\newcommand{\pqx}[1]{{\partial^2  \over \partial x^2}}            
\newcommand{\th}{^{\mit th}}

\def\th{\theta}

\def\t{\tau}

\def\<{\langle}
\def\>{\rangle}

\newfont{\gothic}{eufm10 scaled\magstep1}

\begin{document}
\begin{titlepage}
\begin{flushleft}
\today
\end{flushleft}
\begin{center}
{\LARGE \bf Lightlike infinity in $CGA$ models of Spacetime}
\end{center}
\vskip 2cm
\begin{center}
\mbox{F.M.C.\ Witte}\\
{
\it
Julius School of Physics and Astronomy, 
Utrecht University\\
Leuvenlaan 4, 3584 CE, Utrecht\\
Netherlands}
\end{center}
\vskip 1.5cm
\begin{abstract}
This paper discusses a $7$ dimensional conformal geometric algebra model for spacetime based on the notion that spacelike and timelike infinities are distinct. I show how naturally of the dimensions represents the lightlike infinity and appears redundant in computations, yet usefull in interpretation.
\end{abstract}
\end{titlepage}
\newpage
\newsection{Introduction}
Geometric algebra arises from a fruitfull mariage between vectoralgebra and Clifford algebra. It was Hestenes \cite{STA} who brought it into its presently most common form and who
realised the potential of this mathematical language for physics. Recently textbooks have started appearing \cite{textbooks}. Currently geometric algebra, as applied in physics, comes into two forms, the straightforward direct implementation as found, for example, in spacetime algebra \cite{STA} or the para-vector approach by Baylis \cite{ASP}. A formulation that is especially suited to studying geometric relations has been given by several authors \cite{dorst}, and the same formalism has also been applied to study problems in elasticity theory \cite{elasticity}.

Here three dimensional euclidean space ${\cal{E}}_{3}$ is modelled by means of a five dimensional geometric algebra, $CGA_{5}$. The model is conformal, in the sense that the elements of the algebra 
encoding geometric properties of ${\cal{E}}_{3}$ only need to be determined up to a scale. The great advantage of this way of dealing with ${\cal{E}}_{3}$ is that all conformal transformations
can be treated identically, almost independently of the object to be transformed. For example, the transformation equations for lines, planes, circles, tangents and individual points are the same,
\be
A' = {\cal{R}} A {\tilde{\cal{R}}} \ ,
\ee
where $A$ is a geometric object and ${\cal{R}}$ is a {\it versor}, an element from the algebra that represents the desired transformation.

In relativistic electrodynamics these transformation rules acquire a new use as they can be used to represent the action of electromagnetic fields on proper velocity of a particle \cite{relmech}. The Lorentz force equation 
\be
{\dot{u}(\t)} = F \cdot u(\t) \ ,
\ee
can be turned into a versor-equation for a versor ${\cal{R}}$ in
\be
u(\t) = {\cal{R}} u(0) {\tilde{\cal{R}}} \ ,
\ee
yielding
\be
{\dot{\cal{R}}}  = \frac{1}{2} [{\cal{R}}, F] \ ,
\ee
here we use commutator brackets for shorthand. This technique allows one to solve the equations with considerable ease in a number of interesting cases such as the motion of charges in constant electromagnetic fields, the Coulomb potential and in electromagnetic planewaves \cite{GAem}. Usually it is the position dependence of $F$ that presents problems when one wants to fully exploit the nice
linearity of the equations. The reason is that the versor ${\cal{R}}$ only encodes the evolution of the proper velocity. The evolution of spacetime position must be deduced from that of the velocity, which is only simple when they are uncoupled, or when some symmetry saves the day. If we could extend the versor ${\cal{R}}$ to include the translation as well, we may become able to solve some aspects of this problem.

The versor describing the motion of a charged particle in an electromagnetic field is closely affiliated with the spinor solutions of the Dirac equation \cite{QM}. In fact, quantum mechanics and classical mechanics become very close kin when formulated in terms of versors. In a sense, the Dirac equations remedies the lack of knowledge of the translational degrees of freedom by embedding them in a field theory. If we solve the Dirac equation the spinor at a point carries information on the local direction of a 'proper velocity' connected to the Dirac current, the spinor field contains the information concerning the possible trajectories. The connection
between versors and quantum mechanical wavefunctions, or fermi fields, has not been fully understood yet. A versor description of motion that includes translation will add to this.

In this paper I want to present a $7$-dimensional model of spacetime that does just this. In section 2 I review the $CGA_{5}$ model for $3$-dimensional euclidean space emphasizing those aspects that are important for the discussion in this paper. In section 3 I present
how a $CGA_{7}$ model for spacetime is seen to arise from a proper treatment of the different types of infinity arising in relativity. I also show that it reduces effectively to a $d=4+2$ model by seperating off the light-like infinity that disconnects from the rest 
of the model. The remaining $6$-dimensional conformal model of spacetime has just the properties to ensure that all conformal maps on spacetime are versor operations. This ofcourse includes translations. In section 4 I summarize the main points.

\newsection{The $CGA_{5}$-model for Space}
As was mentioned in the introduction, $3$-dimensional euclidean space can be modelled with great sophistication by a conformal model based on a $5$-dimensional geometric algebra, $CGA_{5}$. I will refer to an excellent tutorial available on-line, in \cite{cga5tut} by Dorst and Fontijne, for more background on this model. Since I will take the $CGA_{5}$ model as a starting point, a brief recapitulation nevertheless seems apropriate.

\newsubsection{Reviewing the $CGA_{5}$ Model}
Euclidean space, ${\cal{E}}_{3}$, can be described as a subspace of higher-dimensional space in a variety of ways. For the $CGA_{5}$ model we start out with a $5$-dimensional vectorspace with
the ordinary vectorspace properties. Next we assume that in this space we have three basis vectors, ${\vec{e}}_{1}$, ${\vec{e}}_{2}$ and ${\vec{e}}_{3}$, in which the position of any point in ${\cal{E}}_{3}$ relative to some origin can be expanded. We will denote the origin by the basis vector ${\vec{O}}$. The last basis vector we need will be denoted ${\vec{e}}_{\infty}$ and its use becomes clear when we define how to compute the euclidean length in this model. Let 
$d_{E}(P,Q)$ denote the euclidean distance between the points $P$ and $Q$. Now suppose that these points are represented by $5$-dimensional vectors ${\vec{p}}$ and ${\vec{q}}$. Then we define, 
\be
- \frac{1}{2} d_{E}(P,Q)^2 = \frac{{\vec{p}} \cdot {\vec{q}}}{({\vec{p}} \cdot {\vec{e}}_{\infty})
({\vec{q}} \cdot {\vec{e}}_{\infty})} \ .
\ee
The dot-product is related in the usual way to the underlying geometric product through
\be
{\vec{p}} {\vec{q}} = {\vec{p}} \cdot {\vec{q}} + {\vec{p}} \wedge {\vec{q}} \ ,
\ee
and represents the symmetric part of the geometric product, whereas the wedge-product represents the antisymmetric part. The fact that the symmetric part yields a scalar is a direct result of the axioms defining any geometric algebra. The model for ${\cal{E}}_{3}$ is found by representing points in ${\cal{E}}_{3}$ by those vectors ${\vec{p}}$ for which,
\be
{\vec{p}} \cdot {\vec{e}}_{\infty} = -1 \ .
\ee
As a points has a vanishing distance to itself, consistency requires a second restriction on the vectors representing points in ${\cal{E}}_{3}$, namely
\be
{\vec{p}}^2 = 0 \ .
\ee
The basis vector ${\vec{e}}_{\infty}$ represents the {\it point at infinity} and also satisfies ${\vec{e}}_{\infty}^{2} = 0$, hence it is at an infinite distance to any other point in ${\cal{E}}_{3}$. We are dealing with an indefinite metric. Let us denote the base vector ${\vec{O}}$ by ${\vec{e}}_{4}$, if we diagonalise the matrix $g_{ij} = {\vec{e}}_{i} \cdot {\vec{e}}_{j}$, that we obtain from the inner products of the matrices, then we find this space has a signature 
$+3$. 
\newsubsection{Versors and vectors is $CGA_{5}$}
The $CGA_{5}$ model allows for subtle distinctions between various vector concepts. One of these is particularly important for our discussion of kinematics in section 4. There we will need
the notion of a {\it tangent vector at a point}. In the $CGA_{5}$ model this distinction arises from the fact that the basis vectors ${\vec{e}_{j}}$ for $j = 1, 2, 3$, can be wedged with the vector representing the origin to designate just such a combination of vector-properties {\it and}an association with a particular point. So, given some {\it pure space vector}
\be
{\vec{v}} = v_{1}{\vec{e}_{1}} + v_{2}{\vec{e}_{2}} + v_{3}{\vec{e}_{3}} \ ,
\ee
we will define the tangent vector $V$ at ${\vec{O}}$ by
\be
V = {\vec{O}} \wedge {\vec{v}} \ .
\ee
It is important to be aware that the tangent vectors at other positions can be found from the one at the origin by applying a translation. Such translations do not only affect ${\vec{O}}$,
but also ${\vec{v}}$. It is relatively straightforward to check that, given another pure space vector ${\vec{x}}$, the versor
\be
T({\vec{x}}) = ( 1 - \frac{1}{2} {\vec{x}} {\vec{e}}_{\infty}) \ ,
\ee
can be used to represent the translated origin ${\vec{O}}'$ as follows
\be
{\vec{O}}' = T({\vec{x}}) {\vec{O}} {\tilde{T}}({\vec{x}}) \ .
\ee
This versor does {\it not} leave pure space vectors invariant!

\newsubsection{Linear Sub-spaces}
The great utillity of this model lies in the ease with which certain classes of subspaces can be represented. We will just mention a few example which are relevant to the discussion in this paper. So-called {\it Flats} like lines, planes and point-pairs, as well as {\it Rounds} such as circles and spheres can be represented by blades, i.e. wedge-products of vectors. For example, the points ${\vec{a}}$, ${\vec{b}}$, ${\vec{c}}$ and ${\vec{d}}$ define a unique sphere passing containing each point. The blade
\be
S = {\vec{a}} \wedge {\vec{b}} \wedge {\vec{c}} \wedge {\vec{d}} \ ,
\ee
describes just this sphere, for any point ${\vec{x}}$ satisfying
\be
{\vec{x}} \wedge S = 0 \ ,
\ee
is on $S$. In this model we can construct geometric objects by wedging points together, with a special role for the vector ${\vec{e}}_{\infty}$; if it is part of a blade, this blade typically describes flats like lines and planes. Even a simple point ${\vec{p}}$ in ${\cal{E}}_{3}$ can be wedged with infinity to yield a so-called {\it flat point} ${\vec{p}} \wedge {\vec{e}}_{\infty}$. Flatpoints are points that are connected to infinity in an intricate sense, as I shall illustrate below.

Geometric objects can be specified in the $CGA_{5}$ language using the wedge-product and points contained int he object. But there are alternatives, and I will mention the one most relevant for this paper. Dualisation allows us to move to an inner-product representation of geometric structures. There exists a dual representation of the sphere $S$, denoted $S^{D}$ for which any point ${\vec{x}}$ on $S$ satisfies
\be
{\vec{x}} \cdot S^{D} = 0 \ .
\ee
This dualisation is done with respect to the {\it smallest common subspace} of the elements contained in the object. In many cases taking the dual amounts to multiplication by the unit-pseudoscalar in the geometric algebra. An important application of the dual representation is in determining the {\it intersection} of two subsets. With a slight abuse of notation the following holds,
\be
(A \cap B )^{D} = A^{D} \wedge B^{D} \ . 
\ee
The sphere is a grade 4 blade and thus the {\it dual sphere} is grade 1, i.e. a vector! This close proximity of spheres and points is no coincidence. It is relatively straightforward to show that, given a dual sphere $S^{D}$, the radius $R$ of the sphere is given by
\be
R^2 = S^{D} \cdot S^{D} \ ,
\ee
ordinary points are just dual spheres of zero radius. The $CGA_{5}$ model also contains spheres with {\it negative} squared radii, referred to as imaginary spheres, that do not have an immediate geometric interpretation despite the fact that they are important elements needed to close the algebra. We will see later that in the $CGA_{7}$ model of spacetime the analogons of imaginary dual spheres are real geometric objects.

I end this review of the $CGA_{5}$ model with one more note on dual spheres, as it nicely illustrate the nature of the flat-point. The dual representation of a sphere around a point ${\vec{q}}$ and containing a point ${\vec{p}}$ can be show to be
\be
S({\vec{p}},{\vec{q}}) = {\vec{p}} \cdot ({\vec{q}} \wedge {\vec{e}}_{\infty}) \ ,
\ee
which can be read as representing {\it the object containing ${\vec{p}}$ and orthogonal
to the flat-point ${\vec{e}}_{\infty}$}. If we picture flat-points as a points with 
{\it hair} extending all the way to infinity, this description is very satisfactorilly.

\newsection{The $CGA_{7}$ Model for Spacetime}
What we should have picked up from the above is that the proper treatment of the point at infinity has been the key to developping a powerfull new language for phrasing geometric relationships. In this section we seek to approach the construction of a spacetime model
from just this point of view. In this subsection we will do this and show how the model effectively reduces to a $CGA_{6}$ model.

In spacetime the metric is non-definite. As a consequence there exist three forms of infinity. As it is customary to denote spacetime points as {\it events} we will use this word in this context to. Let $d_{M}(P,Q)$ denote the Minkowksi-distance between the events $P$ and $Q$. If we consider some arbitrary spacetime event $P$ we can write down two infinities
\ba
d_{M}(P,{\vec{\infty}}_{+})^{2} & = & + \infty \ , \\
d_{M}(P,{\vec{\infty}}_{-})^{2} & = & - \infty \ .
\ea
The first of the two I will refer to as {\it timelike infinity} it is the event representing the future, or past, infinity that can be reached via timelike worldlines only. The second one is the {\it spacelike infinity} that is a close kin of the point at infinity in the $CGA_{5}$ model for ${\cal{E}}_{3}$. These definitions raise the question whether we should not also introduce a {\it lightlike infinity}. The answer is negative as the lightlike infinity will make
its appearance very naturally in the present context. Equipped with two infinities the construction of a proper, conformally invariant, metric becomes a little more complicated. Next to this, the geometric algebra that is to model spacetime needs one dimension more that one naively might have expected.

Let us first solve the problem of defining a proper metric. The problem associated with doing so is located in the denominator of the definition; what to choose? There are a few criteria. First of all, the relations established to define the infinities must ofcourse hold true. A crucial issue is that of what to say about the distance between the two infinities. Next, the innerproducts between the vector ${\vec{x}}$ representing some event $X$, and the vectors ${\vec{\infty}}_{\pm}$ representing the two infinities should stay finite for the whole algebra to make any sense at all. As vectors representing events will come out as null-vectors, i.e. ${\vec{x}}^{2} = 0$, so should the infinities if we want them to represent actual elements of spacetime. Straightforward choices for evidently symmetric definitions of $d_{M}(P,Q)$ tend to give inconsistencies. Here we present a definition that bears a small amount of assymetry,
\be
- \frac{1}{2} d_{M}(P,Q)^2 = \frac{{\vec{p}} \cdot {\vec{q}}}{({\vec{p}} \cdot {\vec{\infty}}_{+})
({\vec{q}} \cdot {\vec{\infty}}_{-})} \ .
\ee
However, together with the requirements
\ba
{\vec{x}} \cdot {\vec{\infty}}_{\pm} & = & \pm 1 \ , \\
{\vec{\infty}}_{\pm} \cdot {\vec{\infty}}_{\pm} & = & 0^{\pm} \ , \\
{\vec{\infty}}_{+} \cdot {\vec{\infty}}_{-} & = & \th \ ,
\ea
this does give a consistent set, and it is completely symmetric on all events whose distance from an arbitrary origin ${\vec{O}}$ remains {\it finite}. In other words, the assymetry is restricted to and directly related with the structure at infinity itself. The vectors representing infinity are null-vectors in a limiting sense. The $0^{\pm}$ indicates that they are taken to be approaching $0$ from the positive/negative side. The inner-product between the two infinities is taken to be a vanishingly small parameter that is set to $0$ at the end of any computation. All of this seems to suggest that we are dealing with a singular situation, which could be due
to over-counting the number of dimensions that we need. If we diagonalise the matrix of the innerproducts of all 7 basis vectors we find it has two positive, 4 negative eigenvalues and one zero-mode. The latter is easilly identified as the eigenvalue belonging to the vector
\be
{\vec{\Omega}}_{0} = \frac{1}{2}({\vec{\infty}}_{+} + {\vec{\infty}}_{-}) \ .
\ee
If we take the innerproduct of this vector with any vector ${\vec{x}}$ representing an event in spacetime, we find that it vanishes. The natural intepretation of ${\vec{\Omega}}_{0}$ is that it represent an infnity at vanishing Minkowski distance, i.e. {\it lightlike infinity}. The linearly independent combination
\be
{\vec{\Omega}}_{\infty} = \frac{1}{2} ({\vec{\infty}}_{-} - {\vec{\infty}}_{+}) \ ,
\ee
represents a {\it new} type of infinity that contains both spacelike as well as timelike infinities and that satisfies
\be
{\vec{x}} \cdot {\vec{\Omega}}_{\infty} = -1 \ ,
\ee
with {\it all} events in spacetime. This linear combination of spacelike and timelike infinity actually takes over the role of ${\vec{\infty}}$ in the $CGA_{5}$ model of ${\cal{E}}_{3}$.

Lightlike infinity completely decouples from the model and therefor the $CGA_{7}$ model effectively reduces to a $CGA_{6}$ model. The eigenvalues of the inner-product matrix indicate that we are dealing with a $SO(2,4)$ invariant metric definition, and it should be no surprise that this group which is known to be the conformal group of Minkowski space \cite{confgr} appears here. If ${\vec{x}}$ is a pure spacetime vector, the versors of the form
\be
V = \exp{(\frac{1}{2}{\vec{x}}{\vec{\Omega}}_{\infty}  )} \ ,
\ee
generate the ordinary spacetime translations. However, minding the fact that there is another infinity in the game, we could also write down versors like
\be
V = \exp{(\frac{1}{2}{\vec{x}}{\vec{\Omega}}_{0}  )} \ .
\ee
Their geometric role needs further clarification\cite{witte2}.

\newsubsection{subspaces}
The spacetime analogon of ${\cal{E}}_{3}$ dual spheres are objects with the following representation,
\be
S({\vec{c}}) = {\vec{c}} + R^2 {\vec{\Omega}}_{\infty}  \ .
\ee
This is the dual representation of an object with a {\it constant Minkowski radius} which I shall refer to generally as shells. If $R^2 \geq 0$, then we are dealing with a timelike shell at some fixed distance from the event ${\vec{c}}$, if $R^2 \leq 0$ we are dealing with a spacelike object that we may want to call a {\it dynamical sphere}. It is a spherical surface collapsing in on the event ${\vec{c}}$ and then expanding again, all at a constant acceleration.
The reader may have felt some surprise that the $CGA_{5}$ model for ${\cal{E}}_{3}$ was capable of representing non-linear objects such as spheres and circles in a linear fashion. Here we see the spacetime version of this; dynamically non-uniform and non-linear elements 
being described by means of a static and linear equation of the form
\be
{\vec{x}} \cdot S^{D} = 0 \ .
\ee
To provide you with a physical systems in which such $3$-shells occur, here are two straightforward examples. 

Suppose that, following a heavy-ion collision taking place at some event ${\vec{c}}$, a quark-gluon plasma hadronizes after a proper-time $\t$ \cite{qgp}. Now two observers measure the radiation coming from the hadronization and they do so at the events ${\vec{p}}_{1}$ and ${\vec{p}}_{2}$. To determine which part of the hadronization process they observe, all we need to do is form the intersections $H_{1}$ and $H_{2}$ of the lightcones of the observers 
with the shell corresponding to the hadronization process. The shell is given by the dual description
\be
S_{hadr}^{D} = {\vec{c}} + \t^2 {\vec{\Omega}}_{\infty} \ ,
\ee
and so we obtain
\be
H_{j} = ({\vec{p}}_{j} \wedge S_{hadr}^{D} )^{D}  \ .
\ee 
This blade completely specifies the subspace in spacetime observed by observer $j$. All the hadronization events, ${\vec{X}}$, observed by observer $j$ are solutions to a linear equation,
\be
{\vec{X}}\cdot ({\vec{p}}_{j} \wedge S_{hadr}^{D} )= 0 \ ,
\ee
and hence can easilly be slotted into a Dirac delta function for evaluation with a source or current distribution. Computing the portion of the hadronization that is observed by {\it both} observers is a similar one-line computation.

The decay of false vacua typically proceeds through the formation of critical bubbles that can be represented as so-called {\it bounce} solutions \cite{bounce} in an underlying field theory. These bounces, at zero temperature, typically are spherically symmetric solutions in 4-dimensional euclidean field theory, as they are imaginary time solutions. The bubble wall separating the different phases typically occurs around some fixed value of the euclidean distance from the centre of the bounce. Whence continued to real time, this bubble wall is a physical realisation of what I have called a dynamic $3$-sphere. Hence it is described by a particular blade in the $CGA_{6}$ algebra. Collisions between bubblewalls occur at the intersections of these blades. Thus the above methods should provide fruitfull in such problems.

\newsection{Discussion and Conclusions}
The six-dimensional $CGA_{6}$ has also been discused previously in \cite{cga6}. I would like to underline once again here that such models could possibly serve as a tool to identify what the geometric {\it primitives} of spacetime really are. In ${\cal{E}}_{3}$ they were the flats and the rounds, but also a whole host of subtly different vector-concepts. No doubt the $CGA_{6}$ model for spacetime will contain a similar, possibly far greater, wealth of geometric objects. 

It is important to note that spacetime events are {\it not} dual spheres with zero radius, rather they are {\it dual lightcones}! This suggests that the "building blocks" of spacetime, the events, and the causal structure of spacetime, lightcones, are related by a duality transformation. In the $CGA_{7}$ model lightlike infinity decouples from the algeba. The question of whether this is also true if quantum aspects of geometry must be taken into account has not been attempted here. However, in view of arguments put forward in \cite{witte3} this could be seen as a hint that particle states in spacetime are unitary representations of a little group of a lightlike vector in a space of 7 or more dimensions.

\end{document}